# LabelECG: A Web-based Tool for Distributed Electrocardiogram Annotation

Zijian Ding[1], Shan Qiu[1], Yutong Guo[2] ,Jianping Lin[3, 4], Li Sun[3], Dapeng Fu[5], Zhen Yang[6], Chengquan Li[4], Yang Yu[7], Long Meng[8], Tingting Lv[4, 9], Dan Li[9] and Ping Zhang[9, 4*]

[1] Department of Electronic Engineering, Tsinghua University, Beijing, China
[2] School of Information and Electronics, Beijing Institute of Technology, Beijing, China
[3] Xinheyidian Co. Ltd, Beijing, China
[4] School of clinical Medicine, Tsinghua University, Beijing, China
[5] Chinese Academy of Sciences Zhong Guan Cun Hospital, Beijing, China
[6] ECG Center, Tianjin Wuqing District People's Hospital, Tianjin, China
[7] The Affiliated Hospital of Qingdao University, Qingdao, China
[8] Shandong Mingjia technology Co., Ltd, Taian, China
[9] Department of Cardiology, Beijing Tsinghua Changgung Hospital, Beijing, China
zhpdoc@126.com

**Abstract.** Electrocardiography plays an essential role in diagnosing and screening cardiovascular diseases in daily healthcare. Deep neural networks have shown the potentials to improve the accuracies of arrhythmia detection based on electrocardiograms (ECGs). However, more ECG records with ground truth are needed to promote the development and progression of deep learning techniques in automatic ECG analysis. Here we propose a web-based tool for ECG viewing and annotating, LabelECG. With the facilitation of unified data management, LabelECG is able to distribute large cohorts of ECGs to dozens of technicians and physicians, who can simultaneously make annotations through web-browsers on PCs, tablets and cell phones. Along with the doctors from four hospitals in China, we applied LabelECG to support the annotations of about 15,000 12-lead resting ECG records in three months. These annotated ECGs have successfully supported the First China ECG intelligent Competition. LabelECG will be freely accessible on the Internet to support similar researches, and will also be upgraded through future works.

## 1      Introduction

Electrocardiography is a common approach to diagnose and screen cardiovascular diseases in clinic. Due to its characteristics including non-invasive, easy-to-operate and economical, it's the most widely adopted clinical detection to diagnose arrhythmia, myocardial ischemia and myocardial infarction [1]. In order to deal with large amounts of electrocardiograms (ECGs), computerized interpretations aim to improve the correctness of ECG diagnose and alleviate the workloads of physicians [2]. Though there are dozens of computerized interpretation systems for ECGs, e.g. GE Marquette system [3] and Glasgow system [4], computerized interpretation are still suffered from the limited diagnostic accuracies [2].

As the fast growth and huge success of deep neural networks in computer vision and natural language processing, etc. [5], these computational techniques are expected to impact the area of precision cardiovascular medicine, including automatic ECG interpretation [6]. Recently Hannun et al published their work on detecting and classifying arrhythmia based on single-lead ECG data [7]. The deep ResNet network achieved better results compared to several technicians, which was trained on almost 91,232 records and tested on 328 records from unique patients. Similarly, Attia et al reported that a neural network, trained on 35,970 ECG-echocardiogram pairs and tested on 52,780 ECG records, can screen for asymptomatic left ventricular dysfunction [8]. These works show that deep neural networks are able to improve the diagnostic accuracies based on large collections of ECG records.

However, almost all ECG databases with careful annotations are small in sample sizes, which might inhibit the application and progression of deep neural networks. The application of an image annotation tool named LabelMe [9] laid the foundation of the well-known ImageNet dataset [10] that flourish the research of deep learning. Similarly, an ECG annotation tool can help build large collections of ECG records with ground truth. As a result, ECG databases with large sample sizes will promote the research of deep learning in the computerized analysis.

In this paper, we present a web-based tool for distributed ECG annotation named LabelECG. Physicians and technicians can annotate ECG records with various time lengths and number of leads, through the web-browsers on desktops, laptops, tablets and cell phones at anytime and anywhere. What's more, ECG datasets are under unified management, and can be accessed by several doctors simultaneously. Four doctors used LabelECG to annotate almost 15,000 12-lead resting ECG records. The resulting database has supported the First China ECG AI Challenge [11]. LabelECG will be accessible online to support similar researches.

## 2   Related Work

ECG annotating often requires expert knowledge and laborious work. Considering the tedious clinical workload of doctors, the manipulations should take up less time and improve efficiency. For example, the tool should provide a convenient way to access such that doctors can use spare time to annotate data. What's more, the tool should be responsible for data management such that doctors can focus on the data annotation. However, most previous tools fail to fulfill these requirements.

Most tools are used off-line (see Table 1). As a result, computers with installed tools and user manuals should be provided for doctors, who have to spend time to learn and use these tools. Among all tools shown in Table 1, WaveformECG [16] is the only web-based tool. However due to unknown reasons, the tool is not accessible currently. For most off-line tools, the problem is that doctors have to be responsible for data management.

The major difference between LabelECG and these previous tools is on two aspects. First of all, LabelECG is a web-based tool for distributed ECG annotation. Several doctors can access LabelECG through web-browsers on desktops, laptops, tablets and even cell phones. Compared to WaveformECG, LabelECG is easier to deploy since it's based on docker. As a result, LabelECG is more suitable when doctors cannot upload data to the Internet and have to make annotations in a local area network. Secondly, LabelECG is responsible for data management. Doctors can ignore the manipulation of data and focus on ECG annotation.

**Table 1.** Tools for ECG annotation.

| Name | Access | Functions |
| --- | --- | --- |
| SigViewer [12] | Off-line | Multi-lead viewing, diagnose annotating |
| EcgEditor [13] | Off-line | Multi-lead viewing, QRS detection, diagnose annotating |
| ECG Viewer [14] | Off-line | QRS detection, diagnose annotating |
| BSS_ECG [15] | Off-line | QRS detection, diagnose annotating |
| WaveformECG [16] | Online | Multi-lead viewing, QRS detection, diagnose annotating |

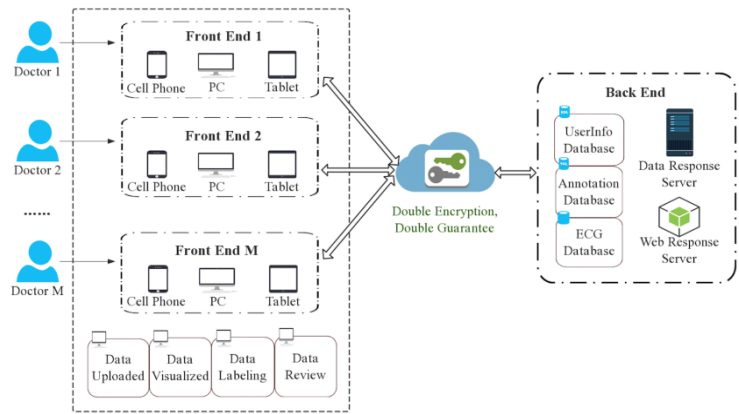

**Fig. 1.** The organization of LabelECG. Multiple doctors can upload, visualize, annotate and review ECG records through the frontends. The backend support these functions through three databases and two servers. LabelECG can be deployed on any cloud systems to connect the frontend and the backend.

## 3   LabelECG

LabelECG is a web-based tool for distributed ECG annotation (see Fig. 1). Through the web browsers of desktops, laptops, tablets and cell phones, multiple doctors can collaborate on annotating the diagnoses of ECG records at any time and any place. With the help of unified data management, doctors are able to focus on annotation without manipulating hundreds of ECG records.

The distributed system consists of a frontend, a backend and a communicating cloud server. To explicitly explain the usage and deployment of LabelECG, we introduce the functions and architecture in the following two sections. Firstly, we mainly discuss how to login LabelECG, choose a dataset, visualize and annotate an ECG record, and revise the personal annotated records. Secondly, we mainly discuss how LabelECG is organized to support the above mentioned functions.

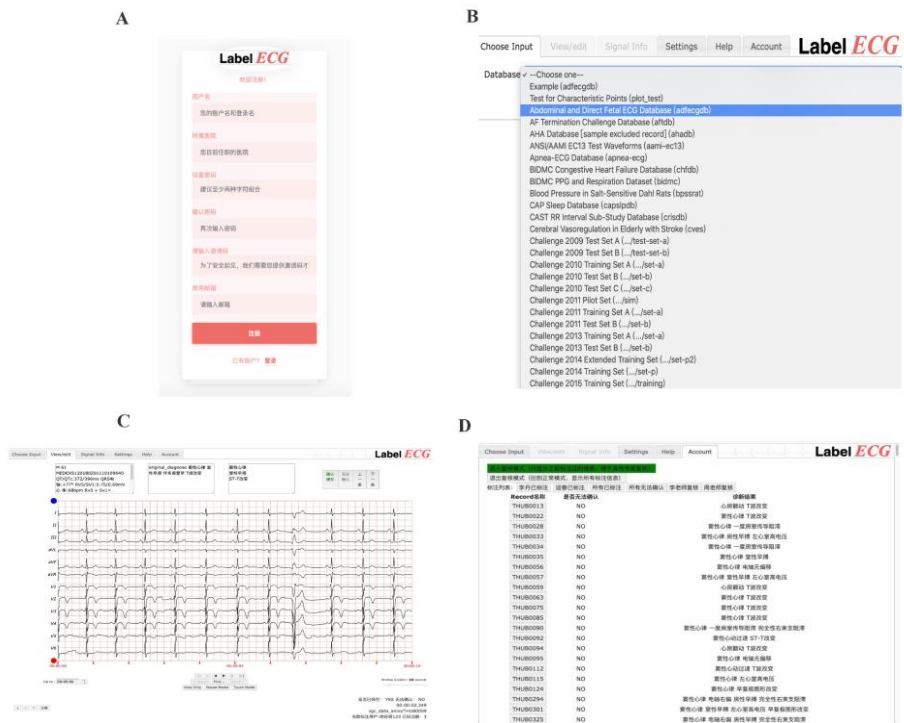

**Fig. 2.** The functions of LabelECG. (A) Register and login; (B) Select a dataset to ECG annotating; (C) Visualize and annotate diagnoses; (D) Personal accounts.

### 3.1 Functions

LabelECG is able to help users upload, visualize, annotate and revise their ECG records. To make these manipulations, the functions of LabelECG is designed to include four parts, including making registration and login, choosing a dataset, annotating the diagnoses, and revising all personal annotations (see Fig. 2).

First of all, the establishment of personal accounts makes it possible to track all annotations of each user. One character is that in consideration of data security, a system administrator needs to provide a verification code to each user to complete registration. This manipulation aims to ensure that only the specific users can have access to their ECG data.

Secondly, LabelECG offers almost all open source datasets from Physionet [17], as well as the user uploaded datasets. Users can choose their own dataset and begin ECG annotating. One characteristic of LabelECG is that when entering the dataset, users can begin with the last record in their last or previous annotation. LabelECG introduces the Lightwave system [17] to visualize any ECG records with various time lengths and number of leads. Meanwhile users can hide specific leads in order to facilitate the observation of certain leads. Three dialog boxes above the visualization are used to help make annotations: the box on the left side provides ECG parameters such heart rate, the one in the middle provides automatic diagnoses, and the one on the right is for writing annotations. Another characteristic of LabelECG is that users can label one record as either "confirmed" or "unsure", since some ECG records may be too ambiguous to annotate.

Furthermore, as multiple users can collaborate on annotating one dataset, LabelECG gives rights to advanced and experienced experts to verify all annotations among these users. Besides, LabelECG enhances intra-group communication by making the unsure ECG data visible for all group members.

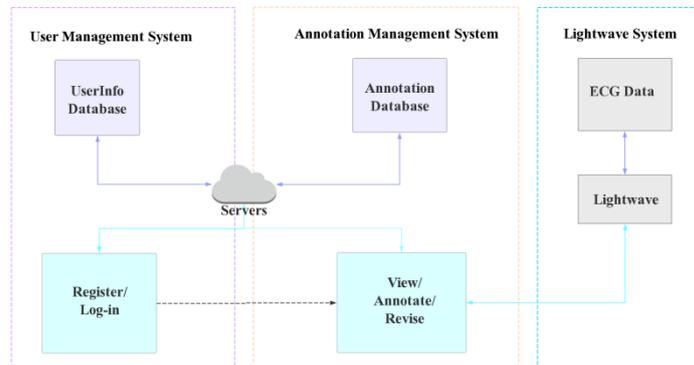

**Fig. 3.** The architecture of LabelECG. A user management system, an annotation management system and the Lightwave system support all the functions of LabelECG.

### 3.2 Architecture

To support the above mentioned functions, LabelECG is built upon three systems, including a user management system, an annotation management system and a Lightwave system (see Fig. 3). The user management system supports the function of user registration and user information management. The annotation management sys-tem supports the function of uploading and storing ECG data and their corresponding parameters, automatic diagnoses and user annotations. The Lightwave system supports the function of ECG visualization.

**User Management System.** This system mainly supports the function of user registration and user information storage. It includes a web page for registration and login, a user information database to store personal accounts, and a server to connect the front web page and the user information database.

To be specific, users need to register their accounts for ECG annotation. We have set "Password" etc. as the required information. After registration, users can login to enter their accounts. The log-in web page sends a login request to the server, and it will check the input information. It will reflect a successful connected prompt if the user's information is found. Otherwise, it will ask the user to check the fill-in information. The frontend of LabelECG offers an online working environment on computers, tablets, or cell phones. The backend server uses the framework of node.js and Express as the Common Gateway Interface (CGI). By keeping abreast with the request from the front end, the back end will return the corresponding data and information.

**Annotation Management System.** This system mainly supports the function of annotating and revising ECG records. It includes an annotation web page, a database to store ECG data and corresponding parameters, automatic diagnoses and user annotations, and a server to connect the frontend annotation web page with the backend database.

To be specific, if users have their ECG data to annotate, there exists an interface to transfer data into the backend ECG signal database. There is a backend server to decode and transfer the raw ECG data into the form fitting the standard of LabelECG. Users are also able to view ECG data from Physionet [17]. After entering the annotation page, the first-time users need to choose the first data to annotate. After the first-time annotation, the system automatically shows the record next to the ones they previously annotated.

In the annotation web page, users can visualize their ECG data and make annotations. We designed a "confirm" button and an "Unsure" button for annotation. After writing annotations into the right side box and pressing the "confirm" button, the server sends this message into the client's list of the Diagnosis Info database. If users press the "Unsure" button, this particular data is stored into a particular list in the Diagnosis Info database. Once users have clicked on one of the mentioned buttons, the interface would automatically turn to the next record. Moreover, we design "Next One" or "Previous One" button for users to view the nearby data.

In order to review the personal annotations, users can press the "Account" button and enter the review pattern. As for regular users, the labeled ECG data and annotation are shown in order. They can click on the data number and enter the annotation page to make revises. If a user has an expert account, s/he can also check other users' annotations.

**Lightwave system.** This system mainly supports the function of ECG data visualization. Physionet provides this system online [17]. We run this system as a CGI application. Once the front end sends requests to the back end, a web server collects and forwards them to the Lightwave system. Afterwards, the Lightwave system will parse the requests and access to the corresponding database in order to obtain data.

## 4      Supporting the First China ECG Intelligent Competition

The First China ECG Intelligent Competition aims to encourage the development of algorithms to classify, from 12-lead resting ECGs with various time lengths, whether an ECG record shows normal, atrial fibrillation, early repolarization and T wave change, etc [11]. In order to ensure the data quality, doctors who come from Beijing Tsinghua Changgung hospital, Chinese Academy of Sciences Zhong Guan Cun Hospital, Tianjin Wuqing District People's Hospital, and the Affiliated Hospital of Qingdao University, used LabelECG to visualize, annotate and review about 15,000 records. LabelECG helped pair four doctors as two teams, and gave rights to two experienced doctors to review and revise all annotated records. With the assistances of LabelECG, these doctors finished annotations in about three months which guaranteed the success of the competition.

## 5      Discussion

LabelECG is a web-based tool for distributed ECG annotation. Multiple doctors can upload, visualize, annotate and revise ECG records via web browsers through desktops, laptops, tablets and cell phones. LabelECG is able to distribute one dataset to several doctors for collaborative annotation. It is also responsible for unified data management such that doctors can focus on data annotation. With doctors as first users, LabelECG supported the First China ECG Intelligent Competition. Our doctors annotated about 15,000 12-lead resting ECG records in about three months.

The current version of LabelECG can make annotations on diagnoses but lacks the ability to annotate local information such as beats and waves. We will add functions including beats annotation and fiducial point annotation, and add more automatic analysis functions, in order to reduce burden and improve efficiency. In addition, since the current version supported the Competition, LabelECG was presented in Chinese. We will distribute this version of LabelECG in English.

## Acknowledgement

This work is supported by The National Key Research and Development Program of China (2017YFB1401804).